\documentclass[preprint]{aastex}

\begin{document}

\title{The sub-Saturn Mass Transiting Planet HAT-P-12b}
\author{Jae Woo Lee, Jae-Hyuck Youn, Seung-Lee Kim, Chung-Uk Lee, and Tobias Cornelius Hinse}
\affil{Korea Astronomy and Space Science Institute, Daejon 305-348, Korea}
\email{jwlee@kasi.re.kr, jhyoon@kasi.re.kr, slkim@kasi.re.kr, leecu@kasi.re.kr, tchinse@gmail.com}

\begin{abstract}
We present new photometric data of the transiting planet HAT-P-12b observed in 2011. Our three transit curves are modelled 
using the JKTEBOP code and adopting the quadratic limb-darkening law. Including our measurements, 18 transit times 
spanning about 4.2 yr were used to determine the improved ephemeris with a transit epoch of 2,454,187.85560$\pm$0.00011 BJD 
and an orbital period of 3.21305961$\pm$0.00000035 d. The physical properties of the star-planet system are computed using 
empirical calibrations from eclipsing binary stars and stellar evolutionary models, combined with both our transit parameters 
and previously-known spectroscopic results. We found that the absolute dimensions of the host star are 
$M_{\rm A}$=0.73$\pm$0.02 M$_\odot$, $R_{\rm A}$=0.70$\pm$0.01 R$_\odot$, log $g_{\rm A}$=4.61$\pm$0.02, 
$\rho_{\rm A}$=2.10$\pm$0.09 $\rho_{\odot}$, and $L_{\rm A}$=0.21$\pm$0.01 L$_\odot$. The planetary companion has 
$M_{\rm b}$=0.21$\pm$0.01 M$_{\rm Jup}$, $R_{\rm b}$=0.94$\pm$0.01 R$_{\rm Jup}$, log $g_{\rm b}$=2.77$\pm$0.02, 
$\rho_{\rm b}$=0.24$\pm$0.01 $\rho_{\rm Jup}$, and $T_{\rm eq}$=960$\pm$14 K. Our results agree well with standard models 
of irradiated gas giants with a core mass of 11.3 M$_{\oplus}$.
\end{abstract}

{}

\section{INTRODUCTION}

Transiting exoplanets are interesting objects to study because the coupling of radial velocity and photometric measurements 
allows a determination of stellar and planetary parameters and thus give us an important constraint on fundamental models of 
planet formation and evolution. For that purpose, we need to precisely measure the physical properties of both the stars and 
planets in these systems from a detailed analysis of transit curves obtained with high-precision photometry. Basically it is 
possible to regard the star-planet systems as detached eclipsing binaries with extremely low mass ratios ($\la$ 0.001). 
Therefore, the methods and tools that are used to model the binary stars could be applied to analyze the light curves of 
transiting exoplanets and then to determine their physical parameters (Southworth 2008, 2009; Lee et al. 2011).

The transiting nature of HAT-P-12b was detected with the HAT-5 telescope of Hungarian-made Automated Telescope Network 
(HATNet; Bakos et al. 2004) in 2006. Hartman et al. (2009) conducted follow-up photometry of four transits together with 
spectroscopic observations and reported that the transit features come from a Saturn-mass planet with a mass of 
0.21$\pm$0.01 M$_{\rm Jup}$ and a radius of 0.96$^{+0.03} _{-0.02}$ R$_{\rm Jup}$ in a 3.2-d circular orbit. 
Its host star is a K4 dwarf GSC 3033-706 (2MASS J13573347+4329367; $V$=$+$12.84) with $M_{\rm A}$=0.73$\pm$0.02 M$_\odot$, 
$R_{\rm A}$=0.70$^{+0.02} _{-0.01}$ R$_\odot$, and [Fe/H]=$-$0.29$\pm$0.05. Since HAT-P-12b is one of 
the lowest-density planets orbiting metal-poor host stars, the physical properties of the system are important for 
irradiation models. In this work, we report and analyze three new high-precision transits of HAT-P-12 and refine 
the physical parameters of the transiting planetary system.

\section{NEW OBSERVATIONS}

We obtained new photometric data from observing three transits of HAT-P-12b. Observations were carried out between 2011 
March and May, using an ARC 4K CCD camera and a Cousins $R$ filter attached to the 1.0-m reflector at 
the Mt. Lemmon Optical Astronomy Observatory (LOAO) in Arizona, USA. The telescope was significantly defocused, 
because it is expected to minimise random and flat fielding errors (see, e.g., Southworth et al. 2009). The e2v CCD chip 
has 4096$\times$4096 pixels and a pixel size of 15 $\mu$m. The image field-of-view is $\sim$ 28$\times$28 arcmin$^2$ at 
the f/7.5 Cassegrain focus of the telescope. A summary of the observations is given in Table 1, where we present 
observing interval, filter, binning mode, exposure time, numbers of observed points, and weather condition. 
During the first run, a gap before the transit ingress was caused by technical problems. With the customary IRAF package, 
we processed the CCD frames to correct for bias level, dark noise, and pixel-to-pixel inhomogeneities of quantum efficiency 
(flat field correction). We applied simple aperture photometry to obtain instrumental magnitudes.

For each transit event we constructed an artificial comparison star by monitoring field stars imaged on the chip. Following 
the method described in Lee et al. (2011), we selected and combined by a weighted average about 10 candidate non-variable 
comparison stars. Then, the differential magnitudes from the artificial reference were normalized by fitting a linear function 
to the out-of-transit data to remove time-varying atmospheric effects (detrending). Resultant transit curves are plotted 
in Figure 1 and listed in Table 2, where times are Barycentric Julian Dates (BJD) in the Barycentric Dynamical Time system 
(Eastman et al. 2010).

\section{LIGHT-CURVE ANALYSIS AND TRANSIT TIMES}

To determine the planetary and orbital parameters, three LOAO transits of HAT-P-12 were analyzed simultaneously in a manner almost 
identical to that for the transiting planetary system TrES-3 (Lee et al. 2011) using the JKTEBOP code (Southworth et al. 2004a,b),
which is a code for modelling the light curves of detached eclipsing binary stars using biaxial spheroids. The main parameters of 
the model are the orbital period ($P$), the ephemeris epoch ($T_0$), the fractional radii of star ($r_{\rm A}$=$R_{\rm A}$/$a$, 
where $a$ is the orbital semi-major axis) and planet ($r_{\rm b}$=$R_{\rm b}$/$a$), the orbital inclination ($i$), and 
the limb-darkening coefficients (LDCs).  Actually, $r_{\rm A}$ and $r_{\rm b}$ are incorporated as their sum ($r_{\rm A}$+$r_{\rm b}$) 
and ratio ($k$=$r_{\rm b}/r_{\rm A}$). Throughout this paper, we refer to star and planet as the subscripts `A' and `b', respectively. 

In the transit analysis, we used $r_{\rm A}$+$r_{\rm b}$ and $k$ as the fitting parameters, because these parameters are more 
weakly correlated than between $r_{\rm A}$ and $r_{\rm b}$ (Southworth 2008). Initial quadratic LDCs were taken from the tables 
of Claret (2000), using the atmospheric parameters of $T_{\rm A}$=4650$\pm$60 K, log $g_{\rm A}$=4.61$\pm$0.01, and 
[Fe/H]=$-$0.29$\pm$0.05 (Hartman et al. 2009). Final results are obtained from fitting the linear LDC ($u_A$) but fixing 
the non-linear LDC ($v_A$), because the two LDCs suffer from strong correlations between them 
(Southworth 2008; Johnson et al. 2008). These are summarized in Table 3, together with the stellar density ($\rho_{\rm A}$) 
and the planetary surface gravity ($g_{\rm b}$) and zero-albedo equilibrium temperature defined as 
$T_{\rm eq}$=$T_{\rm A} \sqrt{r_{\rm A}/2}$ (Southworth 2010). The lower three values were directly calculated from 
those transit parameters and the stellar velocity amplitude ($K_A$=35.8$\pm$1.9 m s$^{-1}$) of Hartman et al. (2009). In order 
to assess the uncertainties of the fitted parameters, we ran 10000 Monte Carlo simulations and a residual permutation algorithm 
(Jenkins et al. 2002) implemented into JKTEBOP, respectively. On these occasions, the non-linear LDC was perturbed by $\pm$0.1 
around the fixed value. The error estimates presented in Table 3 are the 1$\sigma$-values adopted from the larger of the two results. 
Figure 2 displays the light curves with our best-fitting model and residuals.

For ephemeris computations, the minimum times and their uncertainties for our three transits were determined with the JKTEBOP code 
and the model parameters of Table 3. In addition to these, four transit times were newly determined by us from 
the individual measurements of Hartman et al. (2009). The times reported in the discovery paper are not actual events, because 
they were calculated by assuming strictly periodic transits and by adjusting both centers of the first transit and the last transit 
covering all observations in transit modeling. The results are given in Table 4, together with 11 epochs taken from 
Exoplanet Transit Database\footnote {http://var2.astro.cz/ETD/} (ETD; Poddan\'y et al. 2010). The ETD timings were transformed 
from HJD based on UTC into TDB-based BJD using the online applets\footnote {http://astroutils.astronomy.ohio-state.edu/time/} 
developed by Eastman et al. (2010). By including these literature-based transit timings into a linear least-squares fit, 
we computed the orbital ephemeris of HAT-P-12 to be
\begin{equation}
 C_{\rm tr} = \mbox{BJD}~ 2,454,187.85560(11) + 3.21305961(35)E, 
\end{equation}
where $E$ is the number of orbital cycles elapsed from the reference epoch and the parenthesized numbers are 
the 1$\sigma$-error values for the last digit of each term of the ephemeris. Individual mid-transit timing errors were used to 
obtain the corresponding weight factors. The period is somewhat shorter than those of Hartman et al. ($P$=3.2130598$\pm$0.0000021 d) 
and from our transit analysis in Table 3. 

The observed ($O$) {\it minus} calculated ($C_{\rm tr}$) transit times from equation (1) are given in the fourth column of Table 4 
and plotted in Figure 3. The timing residuals seems to indicate a short-term oscillation as is shown in the figure. This might be 
giving a hint about a period change, for example, due to the presence of a second planet in the HAT-P-12 system. 
We examined whether the residuals represent periodic variability, but no detectable periodicity was found. As in the case of 
close binaries with deep convective envelopes (Lee et al. 2009), the apparent transit timing variation in Figure 3 may be produced 
by the sporadic asymmetries of transit curves due to starspots on stellar surface caused by magnetic activity. In reality,
previous measurements (especially from ETD) might have red noise contamination and their errors would then be underestimated
(Carter \& Winn 2009).

\section{RESULTS AND DISCUSSION}

We can compute the absolute demensions of transiting planetary systems by combining photometric and spectroscopic results. 
Currently, there are two main methods to obtain stellar and planetary parameters from the observed quantities. The first method is 
to use the empirical relations from eclipsing binary stars and the second to apply the isochrones from stellar evolutionary models.
For this procedure, we used the planet velocity amplitude $K_{\rm b}$ as a key parameter governing the solution process to find 
the best match between the observations and predictions,  which is the same approach as in Southworth (2009) and Lee et al. (2011). 

First of all, we calculated the physical properties of the HAT-P-12 system using the new calibrations of stellar masses and 
radii expressed as $T_{\rm A}$, log $\rho_{\rm A}$, and [Fe/H] (Enoch et al. 2010, hereafter ECPH), which are originally defined  
by Torres et al. (2010) from 95 well-studied eclipsing binaries and replaced log $g_{\rm A}$ with log $\rho_{\rm A}$. The process 
is to look for $K_{\rm b}$ satisfying simultaneously the stellar mass and radius from the two relations. The calibrated mass 
and radius of $M_{\rm A,ECPH}$= 0.765$\pm$0.012 M$_\odot$ and $R_{\rm A,ECPH}$= 0.686$\pm$0.012 R$_\odot$ were used to minimize 
the $\chi^2$ expressed as:
\begin{equation}
\chi^2 = \biggl[ {{M_{\rm A,ECPH} - M_{\rm A,pred}} \over {\sigma_{M_{\rm A,ECPH}}}} \biggr]^2
+ \biggl[ {{R_{\rm A,ECPH} - R_{\rm A,pred}} \over {\sigma_{R_{\rm A,ECPH}}}} \biggr]^2,
\end{equation}
where $M_{\rm A,pred}$ and $R_{\rm A,pred}$ are the relation-predicted mass and radius from the observations and the $K_{\rm b}$ values. 
The results are given in the second column of Table 5. The quantity $\Theta$ denotes the Safronov (1972) number.

By considering both the metallicities allowed by the observational errors in [Fe/H] and the ages for each metallicity, we use 
three different sets of the stellar evolutionary models: Yonsei-Yale (Demarque et al. 2004), Padova (Girardi et al. 2000), and 
Baraffe et al. (1998, hereafter BCAH). This method also aims to find the velocity amplitude $K_{\rm b}$ for which the measured values 
[Fe/H], $r_{\rm A}$, and $T_{\rm A}$ are best fitted to the radius $R_{\rm A,pred}$ and temperature $T_{\rm A,pred}$ predicted 
from the model isochrones for each metalicity. This consists of calculating the $\chi^2$ fitting statistic,
\begin{equation}
\chi^2 = \biggl[ {{[{\rm Fe/H}] - [{\rm Fe/H}]_{\rm model}} \over {\sigma_{\rm [Fe/H]}}} \biggr]^2
+ \biggl[ {{r_{\rm A} - (R_{\rm A,pred} / a)} \over {\sigma_{r_{\rm A}}}} \biggr]^2
+ \biggl[ {{T_{\rm A} - T_{\rm A,pred}} \over {\sigma_{T_{\rm A}}}} \biggr]^2,
\end{equation}
where $\sigma_{\rm [Fe/H]}$, $\sigma_{r_{\rm A}}$, and $\sigma_{T_{\rm A}}$ are the uncertainties corresponding to the measurements. 
Our process obtained a best-fit model when the BCAH isochrones with [Fe/H]=$-$0.25 were used; both the Yonsei-Yale and 
Padova isochrones cannot constrain the age of the planetary system and favor the metal-rich stellar models.

The isochrones from the BCAH models are plotted in Figure 4 along with the position of HAT-P-12A and the results are listed in 
the third column of Table 5. The mass and radius from BCAH are somewhat smaller than those derived from the empirical calibration 
of eclipsing binary stars, although the physical parameters between the two methods agree with each other within their uncertainties. 
Similar situations are found for WASP-21 (Bouchy et a. 2010; Barros et al. 2011), WASP-37 (Simpson et a. 2011), and 
WASP-39 (Faedi et al. 2011). As is the case for HAT-P-12, the three planet host stars are metal-poor with metalicities 
[Fe/H] of $-$0.46$\pm$0.11, $-$0.40$\pm$0.12, and $-$0.12$\pm$0.10, respectively. On the other hand, 11 eclipsing binaries sampled 
by Torres et al. (2010) are low-metalicity systems ([Fe/H]$<$0.0). Among these only one component star (V636 CenB) is smaller 
than 1 M$_\odot$. From this it follows that the physical properties from the BCAH model for metal-poor low-mass stars seem to be 
more reliable than those from the empirical calibrations. We chose the BCAH solutions as our final results of HAT-P-12.

The location of HAT-P-12b in the mass-radius diagram is shown in Figure 5, together with ten known Saturn-mass transiting exoplanets 
with masses in the range 0.15 M$_{\rm Jup}<M<$ 0.4 M$_{\rm Jup}$, wherein Kepler-16b is a circumbinary transiting planet on 
a nearly circular 229-d orbit around its two parent stars (Doyle et al. 2011). In the same figure, we show constant density contours 
for 0.1, 0.25, 0.5 and 1.0 $\rho_{\rm Jup}$. The physical properties obtained in this study indicate that HAT-P-12b is 
a low-density sub-Saturn mass planet with a mass of 0.21 M$_{\rm Jup}$, a radius of 0.94 R$_{\rm Jup}$, and a mean density of 
$\rho_{\rm b}$=0.24 $\rho_{\rm Jup}$. The results are most similar to those ($M_{\rm b}$=0.20 M$_{\rm Jup}$, 
$R_{\rm b}$=1.00 R$_{\rm Jup}$, $\rho_{\rm b}$=0.19 $\rho_{\rm Jup}$) of the transiting exoplanet HAT-P-18b (Hartman et al. 2011). 
However, HAT-P-12 is younger and more metal-poor than HAT-P-18 with an age of 12.4 Gyr and a metallicity of [Fe/H]=$+$0.10. 
Furthermore, a classification of transiting close-in planets has recently been suggested by Hansen \& Barman (2007). 
With the derived equilibrium temperature of $T_{\rm eq}$=960 K and Safronov number of 0.023 would classify HAT-P-12b to be of 
class II. However, it was recently shown by Southworth (2010) that a correlation between $T_{\rm eq}$ and $\Theta$ appears 
to have little statistical significance. Therefore this classification seems to be of little importance. In addition, 
future observations of secondary transits (or occultations) of HAT-P-12b might prove to be difficult due to 
the planet's low equilibrium temperature.

We compared our mass and radius to the predicted values from theoretical models of Fortney et al. (2007) with various core masses. 
For these, their models were interpolated to the age of 3.2 Gyr and the solar equivalent semi-major axis of 0.0843$\pm$0.0029 AU 
calculated from $a \sqrt{L_{\rm A}}$ and were plotted in Figure 5 as solid curves. We conclude that HAT-P-12b is 
a H/He-dominated gas giant planet with a core mass of 11.3$^{+2.6} _{-2.1}$ M$_{\oplus}$ and is moderately irradiated by 
its low-metallicity host star. Of the 10 circumstellar transiting planets, eight exoplanets including HAT-P-12b follow 
a suggestive correlation between the inferred core mass and host star's metallicity (Guillot et al. 2006; Burrows et al.\ 2007), 
while the recently discovered planets HAT-P-18b and HAT-P-19b (Hartman et al. 2011) with negligible core masses but 
super-solar metallicities disagree with the prediction. Further discoveries will help to identify and understand 
the possible correlation between planetary parameters such as metallicity, core mass, radius, and equilibrium temperature.

\acknowledgments{ }

The authors wish to thank the staffs of LOAO for assistance with our observations. We appreciate the careful reading and 
valuable comments of the anonymous referee. We have used the Simbad Database maintained at CDS and the Exoplanet Transit Database 
in this research. This work was supported by the KASI (Korea Astronomy and Space Science Institute) grant 2012-1-410-02.

\clearpage
\begin{figure}
 \includegraphics[]{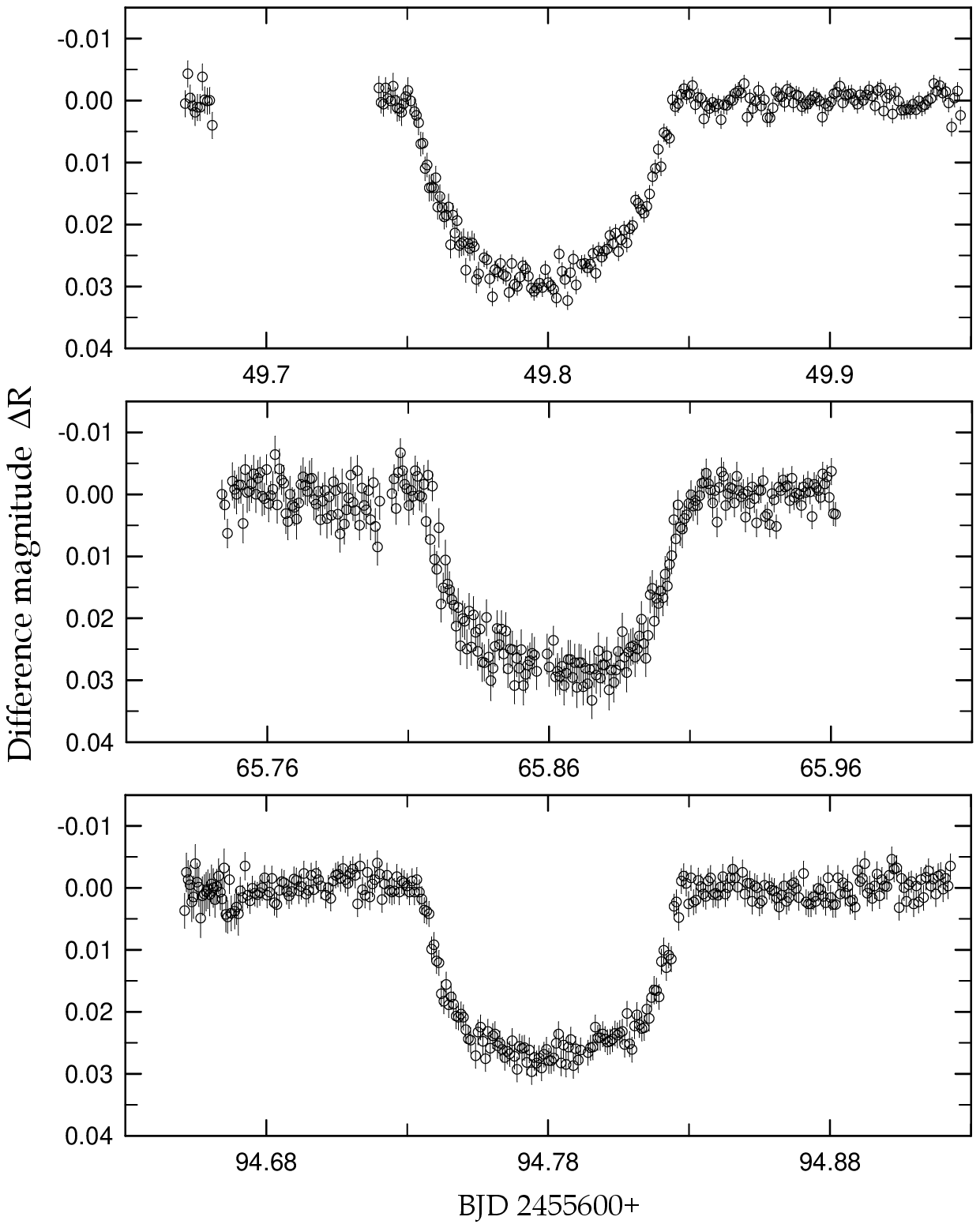}
 \caption{Transit light curves of HAT-P-12 observed between 2011 March and May. }
 \label{Fig1}
\end{figure}

\begin{figure}
 \includegraphics[]{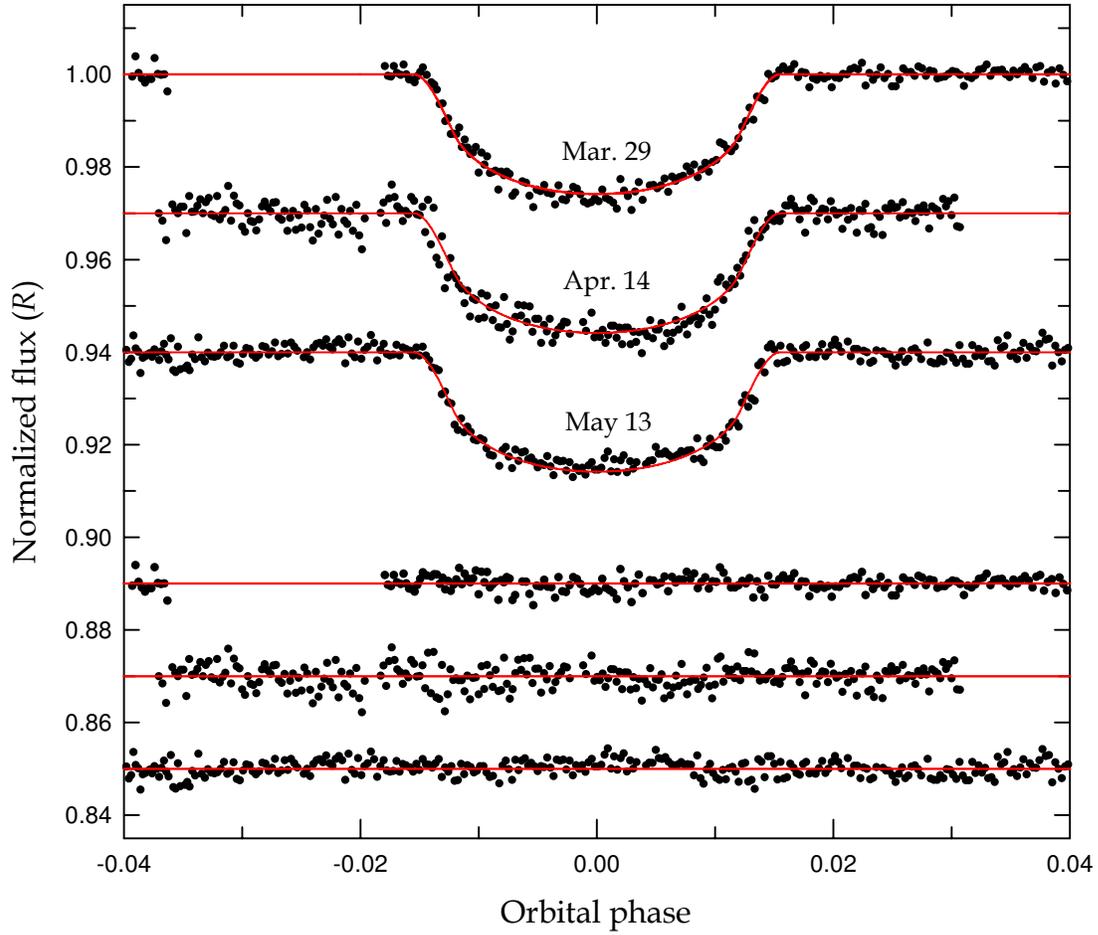}
 \caption{Phased light curves of HAT-P-12. The continuous curves represent the solutions obtained with the best-fit parameters listed 
 in Table 3. The residuals from the fit are offset from zero and plotted at the bottom in the same order as the transit curves. }
 \label{Fig2}
\end{figure}

\begin{figure}
 \includegraphics[]{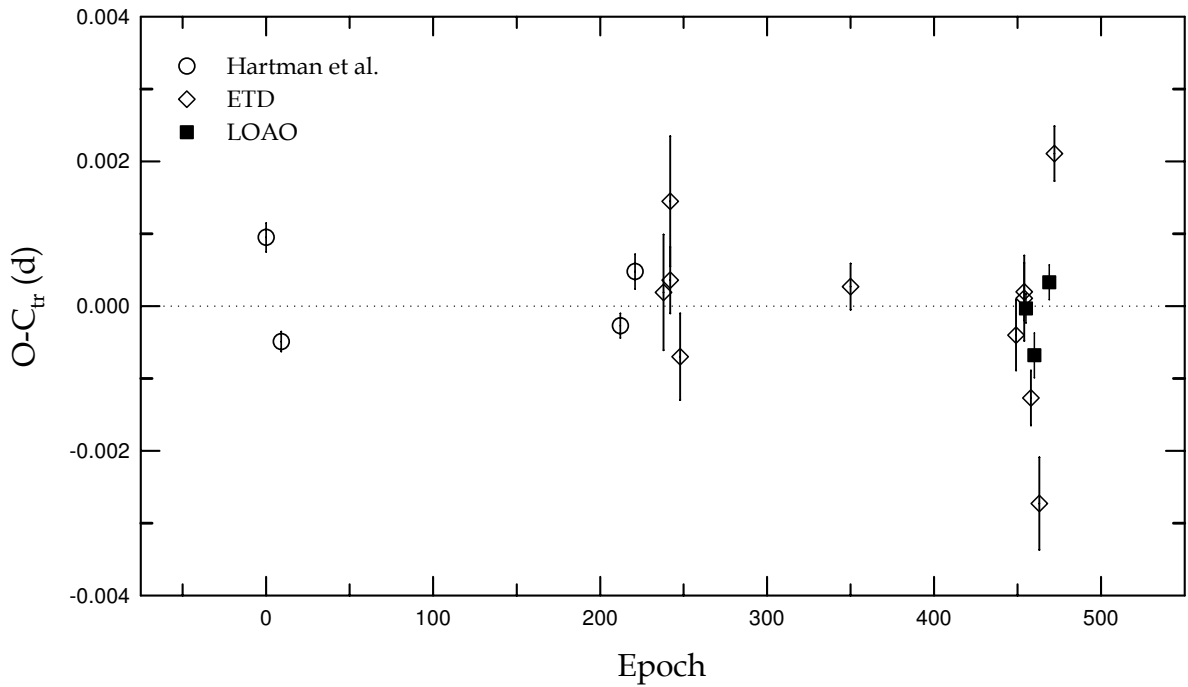}
 \caption{($O-C_{\rm tr}$) residuals for HAT-P-12b constructed with the transit ephemeris (1) newly derived in this work. 
 The minimum times calculated from our transits are plotted using the filled squares. }
 \label{Fig3}
\end{figure}

\begin{figure}
 \includegraphics[]{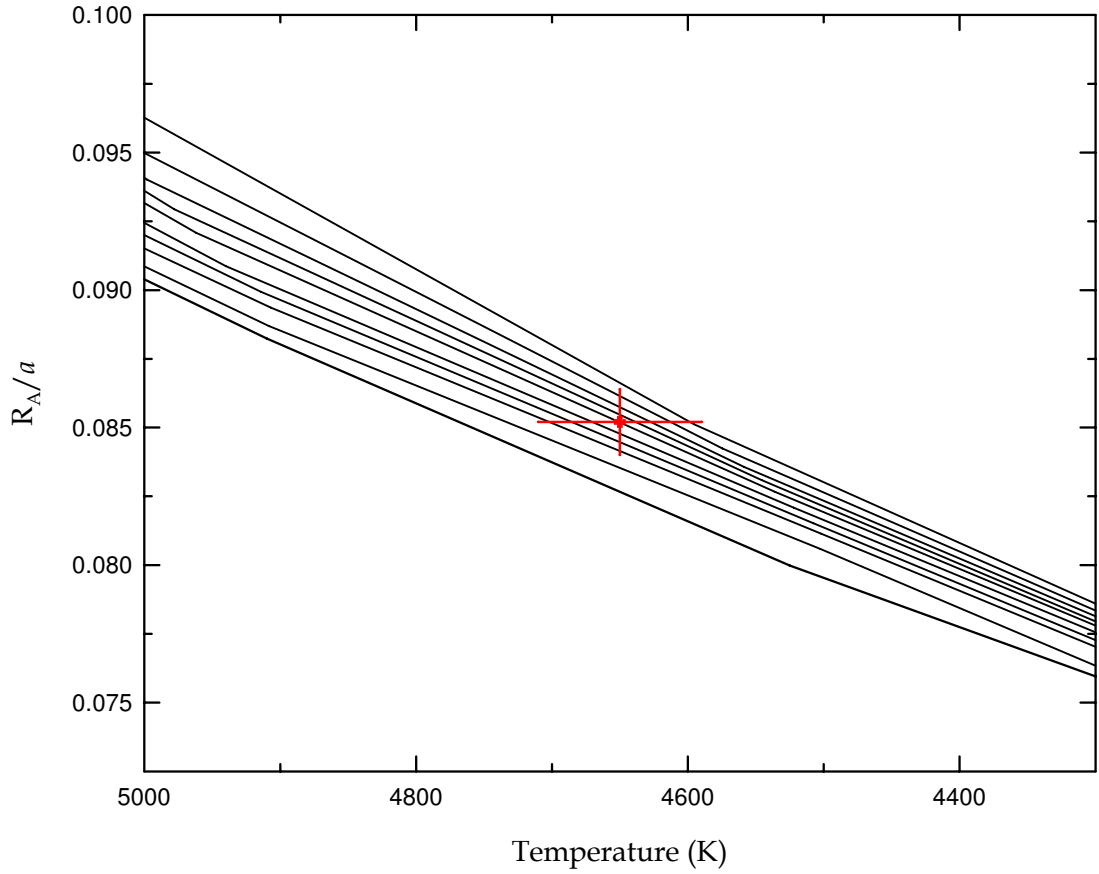}
 \caption{Isochrones from the BCAH models for $\log$ ages = 8.0 (bottom), 8.3, 8.7, 9.0, 9.3, 9.5, 9.6, 9.7, 9.8, and 9.9 yr (top). 
 The observed values of $T_{\rm A}$ and $r_{\rm A}$(=$R_{\rm A}/a$) for HAT-P-12A are shown together with their error bars. }
 \label{Fig4}
\end{figure}

\begin{figure}
 \includegraphics[]{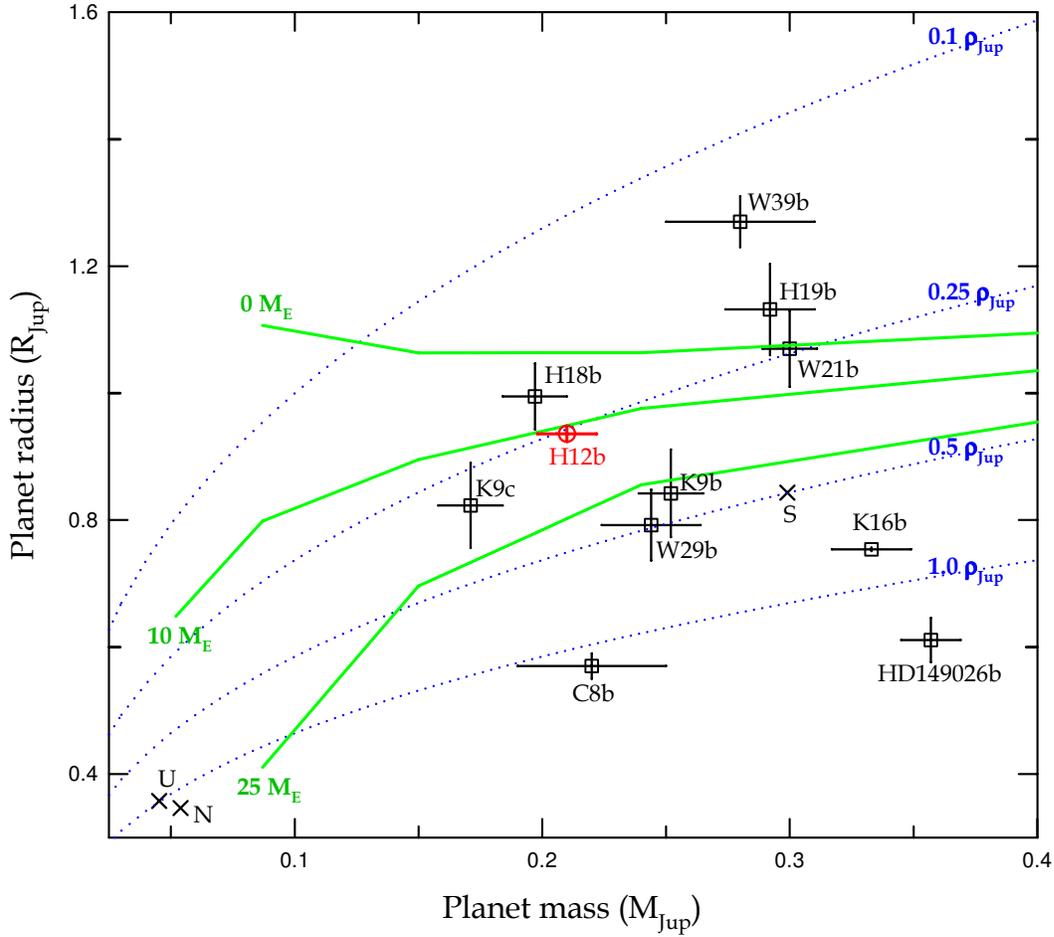}
 \caption{Mass-radius diagram of Saturn-mass transiting planets. W=WASP, H=HAT, K=Kepler, C=CoRoT, and symbols S, N, and U mark 
 the location of Saturn, Nepturn, and Uranus. Dotted lines are density contours for 0.1, 0.25, 0.5, and 1.0 $\rho_{\rm Jup}$.
 Overlaid are Fortney et al. (2007) models interpolated to the solar equivalent semi-major axis of 0.0843 AU and the age of 
 3.2 Gyr for HAT-P-12b with core masses of 0, 10, and 25 M$_{\oplus}$. Data are taken from http://exoplanet.eu/ except for 
 our result for HAT-P-12b.}
 \label{Fig5}
\end{figure}

\clearpage
\begin{deluxetable}{llcccccc}
\tabletypesize{\small}
\tablewidth{0pt} 
\tablecaption{Observing log of HAT-P-12.}
\tablehead{
\colhead{Transit}  &  \colhead{UT Date}  &  \colhead{Observing Interval}  &  \colhead{Filter}   &  \colhead{Binning}  &  \colhead{Exposure}  &  \colhead{$N_{\rm obs}$}  &  \colhead{Weather}    \\
                   &                     &  (BJD+2,455,000)               &                     &  \colhead{Mode}     &  \colhead{Time (s)}  &                           &  \colhead{Condition}  
}
\startdata
1                  &  2011 03 29         &  649.67$-$649.95               &  $R_{\rm C}$        &  2$\times$2         &  50$-$70             &  224                      &  clear                \\  
2                  &  2011 04 14         &  665.74$-$665.97               &  $R_{\rm C}$        &  2$\times$2         &  50$-$70             &  273                      &  partly cloudy        \\  
3                  &  2011 05 13         &  694.65$-$694.93               &  $R_{\rm C}$        &  2$\times$2         &  40$-$60             &  314                      &  clear                \\  
\enddata
\end{deluxetable}

\begin{deluxetable}{cccc}
\tablewidth{0pt} 
\tablecaption{$R_{\rm C}$-band photometry of HAT-P-12}
\tablehead{
\colhead{BJD} & \colhead{Diff. Mag} & \colhead{$\sigma_{\rm Mag}$} & \colhead{Relative Flux} 
}                                                               
\startdata                                                      
2455649.67120   &  $+$0.0005      &  0.0021   &  0.9995   \\ 
2455649.67207   &  $-$0.0043      &  0.0021   &  1.0039   \\ 
2455649.67293   &  $-$0.0004      &  0.0021   &  1.0003   \\ 
2455649.67380   &  $+$0.0009      &  0.0021   &  0.9991   \\ 
2455649.67467   &  $+$0.0019      &  0.0021   &  0.9982   \\ 
2455649.67554   &  $+$0.0011      &  0.0021   &  0.9990   \\ 
2455649.67641   &  $+$0.0011      &  0.0021   &  0.9990   \\ 
2455649.67729   &  $-$0.0038      &  0.0021   &  1.0035   \\ 
2455649.67815   &  $-$0.0001      &  0.0021   &  1.0001   \\ 
2455649.67902   &  $+$0.0001      &  0.0021   &  0.9999   \\ 
\enddata
\tablecomments{This table is available in its entirety in a machine-readable form in the online journal. 
A portion is shown here for guidance regarding its form and content.}
\end{deluxetable}

\begin{deluxetable}{lc}
\tablewidth{0pt}
\tablecaption{Transit parameters of HAT-P-12.}
\tablehead{
\colhead{Parameter}           & \colhead{Value}
} 
\startdata 
$r_{\rm A} + r_{\rm b}$       &  0.0969$\pm$0.0012              \\
$k$ (=$r_{\rm b}/r_{\rm A}$)  &  0.1370$\pm$0.0019              \\
$i$ (deg)                     &  89.915$\pm$0.098               \\
$u_A$                         &  0.739$\pm$0.069                \\
$v_A$                         &  0.177 perturbed                \\
$T_0$ (BJD)$^a$               &  649.79751$\pm$0.00036          \\
$P$ (d)                       &  3.213089$\pm$0.000037          \\
$r_{\rm A}$                   &  0.0852$\pm$0.0012              \\
$r_{\rm b}$                   &  0.01168$\pm$0.00005            \\
$\sigma$ (mmag)$^b$           &  2.1269                         \\
$\chi^2 _{\rm red}$           &  0.9920                         \\[1.5mm]
$\rho_{\rm A}$ $(\rho_\odot)$ &  2.100$\pm$0.089                \\
$g_{\rm b}$ ($g_{\rm Jup}$)   &  0.257$\pm$0.014                \\
$T_{\rm eq}$ (K)              &  960$\pm$14                     \\
\enddata
\tablenotetext{a}{BJD 2,455,000 is suppressed.}
\tablenotetext{b}{rms scatter of residuals.}
\end{deluxetable}

\clearpage
\begin{deluxetable}{lcrrl}
\tablewidth{0pt}
\tablecaption{Observed transit times of HAT-P-12.}
\tablehead{
\colhead{BJD} & \colhead{Uncertainty} & \colhead{$E$} & \colhead{$O$--$C_{\rm tr}$} & References$\rm^a$               \\
\colhead{(2,450,000+)} & & & & }
\startdata
4,187.85655$\rm^b$  & $\pm$0.00020    &    0          &      0.00095                 &  Hartman et al. (2009)         \\
4,216.77265$\rm^b$  & $\pm$0.00014    &    9          &   $-$0.00049                 &  Hartman et al. (2009)         \\
4,869.02397$\rm^b$  & $\pm$0.00017    &  212          &   $-$0.00027                 &  Hartman et al. (2009)         \\
4,897.94225$\rm^b$  & $\pm$0.00024    &  221          &      0.00048                 &  Hartman et al. (2009)         \\
4,952.56398         & $\pm$0.00080    &  238          &      0.00019                 &  Gregorio (AXA)                \\
4,965.41639         & $\pm$0.00046    &  242          &      0.00036                 &  Kucakova (TRESCA)             \\
4,965.41748         & $\pm$0.00090    &  242          &      0.00145                 &  Ayoimanitis (AXA)             \\
4,984.69368         & $\pm$0.00060    &  248          &   $-$0.00070                 &  Gary (AXA)                    \\
5,312.42673         & $\pm$0.00032    &  350          &      0.00027                 &  Vilagi \& Gajdos (TRESCA)     \\
5,630.51896         & $\pm$0.00049    &  449          &   $-$0.00040                 &  Ivanov \& Sokov (TRESCA)      \\
5,646.58477         & $\pm$0.00059    &  454          &      0.00011                 &  Nicolas (TRESCA)              \\
5,646.58486         & $\pm$0.00040    &  454          &      0.00020                 &  Gajdos \& Vilagi (TRESCA)     \\
5,649.79769         & $\pm$0.00020    &  455          &   $-$0.00003                 &  This paper (LOAO)             \\
5,659.43563         & $\pm$0.00038    &  458          &   $-$0.00127                 &  Ruiz (TRESCA)                 \\
5,665.86234         & $\pm$0.00031    &  460          &   $-$0.00068                 &  This paper (LOAO)             \\
5,675.49947         & $\pm$0.00064    &  463          &   $-$0.00273                 &  Zinoviev \& Sokov (TRESCA)    \\
5,694.78089         & $\pm$0.00024    &  469          &      0.00033                 &  This paper (LOAO)             \\
5,704.42185         & $\pm$0.00038    &  472          &      0.00211                 &  Salas (TRESCA)                \\
\enddata
\tablenotetext{a}{AXA (Amateur eXoplanet Archieve), TRESCA (TRansiting ExoplanetS and CAndidates).}
\tablenotetext{b}{Newly determined by us from the individual measurements.}
\end{deluxetable}

\clearpage
\begin{deluxetable}{lccc}
\tablewidth{0pt} 
\tablecaption{Physical properties of the HAT-P-12 system.}
\tablehead{
\colhead{Parameter}                & \colhead{ECPH Model}         & \colhead{BCAH Model}
}                                  
\startdata                         
$K_{\rm b}$ (km s$^{-1}$)          &  131.5$\pm$3.0              &  129.7$\pm$1.5              \\
$M_{\rm A}$ (M$_\odot$)            &  0.757$\pm$0.038            &  0.727$\pm$0.019            \\
$R_{\rm A}$ (R$_\odot$)            &  0.711$\pm$0.019            &  0.702$\pm$0.013            \\
log $g_{\rm A}$ (cgs)              &  4.613$\pm$0.032            &  4.607$\pm$0.020            \\
$L_{\rm A}$ (L$_\odot$)            &  0.212$\pm$0.016            &  0.206$\pm$0.013            \\
$M_{\rm bol,A}$ (mag)              &  6.434$\pm$0.080            &  6.463$\pm$0.069            \\
$M_{\rm b}$ (M$_{\rm Jup}$)        &  0.216$\pm$0.015            &  0.210$\pm$0.012            \\
$R_{\rm b}$ (R$_{\rm Jup}$)        &  0.949$\pm$0.022            &  0.936$\pm$0.012            \\
$\rho_{\rm b}$ $(\rho_{\rm Jup})$  &  0.236$\pm$0.016            &  0.240$\pm$0.012            \\
$\Theta$                           &  0.0233$\pm$0.0020          &  0.0236$\pm$0.0015          \\
$a$ (AU)                           &  0.03887$\pm$0.00088        &  0.03829$\pm$0.00046        \\
Age (Gyr)                          &                             &  3.2$\pm$3.8                \\
\enddata
\end{deluxetable}

\end{document}